\documentclass[conference]{IEEEtran}

\IEEEoverridecommandlockouts
\usepackage[T1]{fontenc}
\usepackage[utf8]{inputenc}
\usepackage[english]{babel}
\usepackage{microtype}
\usepackage{amsmath, amssymb, upgreek}
\usepackage{booktabs, multirow}
\usepackage{circledsteps}

\usepackage{amsthm}
\newtheorem{definition}{Definition}
\newtheorem{proposition}{Proposition}

\usepackage[hidelinks]{hyperref}

\usepackage{tikz}
\usetikzlibrary{arrows.meta}

\pgfkeys{/csteps/inner color=white}
\pgfkeys{/csteps/fill color=black}

\pgfarrowsdeclare{dot}{dot}
{
  \pgfarrowsleftextend{0pt}
  \pgfarrowsrightextend{0pt}
}
{
  \pgfpathcircle{\pgfqpoint{0pt}{0pt}}{2pt}
  \pgfusepathqfill
}

\newcommand{\Tc}[1]{\mathsf{T}_{#1}}
\newcommand{\DTc}[1]{\Updelta \Tc{#1}}

\begin{document}

\title{One-Time Certificates for Reliable and Secure Document Signing \\
\thanks{This study was financed in part by the Coordenação de Aperfeiçoamento de Pessoal de Nível Superior - Brasil (CAPES) - Finance Code 001.}
}

\author{\IEEEauthorblockN{1\textsuperscript{st} Lucas Mayr}
\IEEEauthorblockA{\textit{Departamento de Informática e Estatística} \\
\textit{Universidade Federal de Santa Catarina}\\
Florianópolis, Brazil \\
lucas.mayr@posgrad.ufsc.br}
\and
\IEEEauthorblockN{2\textsuperscript{nd} Gustavo Zambonin}
\IEEEauthorblockA{\textit{Departamento de Informática e Estatística} \\
\textit{Universidade Federal de Santa Catarina}\\
Florianópolis, Brazil \\
gustavo.zambonin@posgrad.ufsc.br}
\and
\IEEEauthorblockN{3\textsuperscript{rd} Frederico Schardong}
\IEEEauthorblockA{\textit{Departamento de Informática e Estatística} \\
\textit{Universidade Federal de Santa Catarina}\\
Florianópolis, Brazil \\
frederico.schardong@rolante.ifrs.edu.br}
\and
\IEEEauthorblockN{4\textsuperscript{th} Ricardo Custódio}
\IEEEauthorblockA{\textit{Departamento de Informática e Estatística} \\
\textit{Universidade Federal de Santa Catarina}\\
Florianópolis, Brazil \\
ricardo.custodio@ufsc.br}
}

%
%
  
\maketitle

\begin{abstract}
  Electronic documents are signed using private keys and verified using the corresponding digital certificates through the well-known public key infrastructure model. Private keys must be kept in a safe container so they can be reused. This makes private key management a critical component of public key infrastructures with no failproof answer. Therefore, existing solutions must employ cumbersome and often expensive revocation methods to handle private key compromises. We propose a new cryptographic key management model built with long-term, irrevocable digital certificates, each bound to a single document. Our model issues a unique digital certificate for each new document to be signed. We demonstrate that private keys associated with these certificates should be deleted after each signature, eliminating the need to store those keys. Furthermore, we show that these certificates do not require any revocation mechanism to be trusted. We analyze the overhead caused by the frequent generation of new key pairs for each document, provide a security overview and show the advantages over the traditional model.
\end{abstract}



\section{Introduction}


Trusting cryptographic key management is one of the main challenges of the \emph{public key infrastructure} (PKI) model. Private keys are protected by cryptographic hardware or software containers, which have several access control mechanisms. Security concerns are amplified when dealing with laypeople. Traditional models require users to understand the responsibilities of possessing a private key and the expected course of action if it is compromised. As a result, signers use simpler, relatively low-cost cryptographic devices such as smartcards or tokens. In addition, cloud-based key management services eliminate the need for users to carry a device, but using such providers raises privacy concerns~\cite{Mushtaq:201710}.

Another recurring challenge in PKIs is the revocation of digital certificates. Revocation is necessary to account for outdated certificate attributes and compromised private keys. There is no failproof solution to certificate revocation~\cite{Arnes:200002}. Due to these complications, we argue that traditional PKIs are not inclusive, both from a usability and knowledge point of view. On the usability side, it requires users to understand how the PKI model works, its inherent complexities, the importance of good key management, and revocation processes. For laypeople, this is an unreasonable requirement; all citizens should be able to sign electronic documents at their convenience.

\subsubsection*{Contributions}

We propose a new type of digital certificate that yields a more inclusive and user-friendly PKI. We focus on the issuance and revocation of certificates and the creation and validation of digitally signed documents. We propose irrevocable, unique certificates called One-Time Certificates (OTCs). We show that OTCs solve the problem of key management, revocation, and usability at the signer level. We compare our proposal with the traditional public key infrastructure model, evaluating performance and security.

\subsubsection*{Related works}

As previously mentioned, private key management and revocation account for great complexity in the PKI model. It raises costs and is therefore considered the main cause that prevents the widespread use of PKI in communication networks~\cite{Ponemon:2021} and in the most diverse applications that could benefit from its services. This has motivated the search for ways to simplify, if not all, then at least parts of the PKI. We briefly discuss works related to simplifying PKI mechanisms and how they affect usability and technology adoption.

Rivest and Topalovic et al. introduced short-lived digital certificates~\cite{Rivest:199802,Topalovic:201205} to remove the need for revocation at the signer level, issuing them with a very short validity period. They argue that if the certificate validity period were less than the latency of traditional revocation systems, then there would be no reason to revoke certificates. One of the main criticisms of this proposal is the increase in complexity on the side of the certificate holder. Furthermore, \emph{certificate authorities} (CAs) should always be available to issue new certificates on demand.

Simple Public Key Infrastructure~\cite{Rivest:199802} (SPKI) and Simple Distributed Security Infrastructure~\cite{Rivest:199610} (SDSI) are other instances of revocation-less PKIs. These models associate a certificate with a recency level. Thus, an entity may request a certificate with a particular ``freshness'', i.e., a certificate guaranteed to be issued after a certain moment. While the models do not allow certificate revocation, long-term certificates still have the same problems as in a traditional PKI~\cite{Arnes:200002}.

The Automatic Certificate Management Environment protocol~\cite{Barnes:201903} is responsible for the increase in the adoption rate of the TLS protocol~\cite{Aas:201911} through simplification and automation of the issuance of HTTPS certificates. By 2020, approximately 81\% of HTTPS websites were using Let's Encrypt certificates~\cite{Aas:202002}. Its success can also be attributed to the free issuance of the certificates. This effort is mentioned to exemplify how high-security standards can be rapidly embraced due to changes in a previously convoluted process.

The ``certificateless'' proposal~\cite{Al-Riyami:200311} radically changes how PKIs are built; certificates are no longer used to transport keys and attributes. Other derived models~\cite{Gutmann:200208,Gutmann:200402,Scheibelhofer:200504,Topalovic:201205} introduce several trade-offs in exchange for features such as the absence of revocation. Particularly, there is an increase in the complexity of cryptographic key management in these models, among numerous other challenges. Furthermore, these proposals are not interoperable and backward-compatible with existing computer systems. 

We note that there is a particular focus on solving the issue of revocation and that the private key is assumed to be operated by a knowledgeable holder. Furthermore, such issues are primarily approached from the perspective of the TLS protocol. We argue that there is a distinct lack of research from the perspective of the general public when signing and verifying digitally signed documents. We aim to solve this issue via the proposal of the OTC model, which enables secure, long-term, and revocation-less digital certificates. In our model, the underlying PKI is made completely transparent to the end user.



\section{Background}\label{sec:background}

In this section, we present the cryptographic primitives, artifacts, and services needed to understand our proposal. We assume the reader to be comfortable with the general aspects of digital signature schemes and PKIs.

\subsubsection*{Digital certificate lifetime}

\begin{figure}[htbp]
  \centering
  \begin{tikzpicture}[very thick, node distance = 4cm] 
    \draw[dotted, -latex, thin] (0, 0) -- (0.8\linewidth, 0);
    \draw[latex-latex, thin] (0.15\linewidth, 15pt) -- (0.5\linewidth, 15pt);
    \draw[latex-latex, thin] (0.15\linewidth, 30pt) -- (0.7\linewidth, 30pt);
    \draw[dot-|] (0.15\linewidth, 0) -- (0.7\linewidth, 0);
    
    \draw[thick] (0.8\linewidth, -4.5pt) node[below, align=center] {Time};
    \draw[thick] (0.325\linewidth, 14pt) node[above, align=center] {$\DTc{c}$};
    \draw[thick] (0.425\linewidth, 29pt) node[above, align=center] {$\DTc{v}$};
    \draw[thick] (0.05\linewidth, -5pt) node[below, align=center] {$\Tc{i}$};
    \draw[thick] (0.15\linewidth, -5pt) node[below, align=center] {$\Tc{b}$};
    \draw[thick] (0.5\linewidth, -5pt) node[below, align=center] {$\Tc{r}$};
    \draw[thick] (0.7\linewidth, -5pt) node[below, align=center] {$\Tc{a}$};
    
    \draw [thick] (0.05\linewidth, 0) --(0.05\linewidth, -5pt);
    \draw [thick] (0.15\linewidth, 0) -- (0.15\linewidth, -5pt);
    \draw [thick] (0.5\linewidth, 0) -- (0.5\linewidth, -5pt);
    \draw [dotted, thick] (0.15\linewidth, 0) -- (0.15\linewidth, 30pt);
    \draw [dotted, thick] (0.5\linewidth, 0) -- (0.5\linewidth, 15pt);
    \draw [dotted, thick] (0.7\linewidth, 0) -- (0.7\linewidth, 30pt); 
  \end{tikzpicture}
  \caption{The various time references for a digital certificate.}\label{fig:reference_time}
\end{figure}
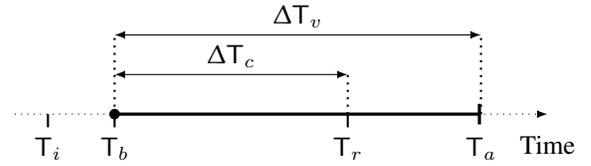

Documents are signed using a private key and verified using the corresponding public-key certificate; certificates cannot be used to verify any signatures after their expiration. A certificate is issued at time $\Tc{i}$ and has a validity period $\DTc{v}$ that begins at $\Tc{b}$ and ends at $\Tc{a}$. A certificate that has been revoked has its validity period reduced to $\DTc{c}$, which is equal to $\Tc{r} - \Tc{b}$, where $\Tc{r}$ is the time of revocation. Such reference times are summarized in Figure~\ref{fig:reference_time}. We observe that, for every certificate, $\Tc{i} \leq \Tc{b} \leq \Tc{r} \leq \Tc{a}$; however, certificates typically follow $\Tc{i} \leq \Tc{b} < \Tc{r} < \Tc{a}$, as issuing an expired or revoked certificate has little to no practical use.

\subsubsection*{Private key containers}

These are specialized structures designed to store, generate, and process private keys safely. Protection levels, authentication methods offered, and the underlying processing power vary wildly. Such features directly reflect on the price of the equipment. Entities that require high levels of security use devices such as \emph{hardware security modules} (HSMs). They offer the most robust physical and logical security levels, often having multiple tamper-resistant features and highly secure cryptographic algorithm implementations. Trusted platform modules (TPMs) are similar hardware-based devices that hold cryptographic secrets, shipped in consumer-grade computer processors. 

Commonly, end users hold \emph{smartcards} and/or \emph{cryptographic tokens}, cheaper hardware alternatives with limited processing power and storage. These devices come in the form of plastic cards or USB sticks, featuring a microprocessor with embedded memory and specialized cryptographic functions. They are widely used in the traditional PKI model~\cite{Longo:200204}. \emph{Key management as a service} (KMaaS) is an alternative method of managing key pairs via cloud-based solutions that have become popular~\cite{Lerman:201207,Kuzminykh:202008} particularly because it does not require users to handle any hardware directly~\cite{Ponemon:2021}.

\subsubsection*{Signature policy}

A policy is a set of technical and orderly criteria by which signatures are created and verified, each according to a particular set of requirements. When signing, the certificate holder selects a policy that matches the requirements established by the application that will consume the signed documents. For example, if a user wants to sign a short-lived non-critical resource for their company, they can choose a basic policy so that fewer artifacts will be needed to create a complete signature. However, suppose a document needs to be archived for future use. In that case, a policy must be chosen that foresees the incorporation of all the artifacts that the recipient will need to validate the signature.

Each policy establishes attributes that must be included in the signature. Some attributes must be signed by the end user, while others can be added by any party afterward, such as timestamps. Policies can be either implicitly or explicitly defined and are usually specified for several forms of electronic document representation. The most used formats are ASN.1, PDF, XML, and JSON. These are semantically equivalent, and most changes occur in the representation of the electronic document in the underlying language~\cite{ETSI-EN-3191021-131:202111}. 

\subsubsection*{Advanced Electronic Signatures}\label{sec:ades}

In regulatory terms, electronic signatures that follow signature policies are called Advanced Electronic Signatures (AdES)~\cite{Kutylowski:202301}. Hereafter, we consider advanced signatures as defined in ETSI EN 319 102-1~\cite{ETSI-EN-3191021-131:202111}. We argue that this set of European standards is far-reaching and robust enough to allow for mature comparisons with our proposal. The standards cover various use cases due to the needs of several countries and allow for the customization and interoperability of various signature formats. We compare our proposal against the CAdES~\cite{ETSI-EN-3191221-121:202110} format, as the features of the ASN.1 language allow for straightforward parallels with other formats without loss of generality.

To generate an advanced electronic signature in the PKI model, it is not enough to simply create an individual signature $\sigma$. A certificate holder may include references to its certificate and those of its chain, as well as revocation artifacts and, optionally, timestamps. Furthermore, other end users can also include their own signed artifacts. Therefore, the signature verification process must be extended to include every such artifact in the final document. That is, a signed document with a timestamp must successfully validate both signatures for the document to be valid, i.e., it must verify the document, the timestamp, and both certification chains and corresponding revocation data. The total cost of verifying an advanced electronic signature is calculated below.

Let $t \in \mathbb{N}$ be the number of timestamps; for $1 \leq i \leq t + 1$, let $c_i \in \mathbb{N}$ be the depth of the certificate chain of a signing certificate, i.e., the number of certificates from the end user to the root CA certificate, and $r_{i} \in \mathbb{N}$ the number of artifacts with revocation data. Then, the total number of signatures $s \in \mathbb{N}$ to be verified in any signed document is $s = \sum_{i = 1}^{t + 1} (1 + c_i + r_i)$.

Typically, certificate chains are three or four certificates deep. Note that every certificate in the chain is a CA, except for the end-user certificate. Therefore, the number of CAs in a chain is $c - 1$. Every CA must also issue its revocation artifact such that $r = c - 1$. We hereafter assume the worst-case scenario for depth-4 and depth-3 certificate chains, i.e., every chain in the signature has the same length. Thus, the aforementioned equation is simplified to $s = 2 c (t + 1)$.



\subsection{PKI challenges}\label{sec:issues}

Revocation and key management are unavoidable requirements of PKIs due to how public key cryptography is designed. The lifetime of the private key is set by the validity period of the certificate. However, if there is any security issue with the private key or its container, or if any of the attributes of the respective certificate become outdated, it is necessary to revoke the certificate. Therefore, it is clear that effective, rapid, and straightforward revocation is not only desirable but necessary. Before presenting our proposal, we briefly discuss these two major factors that contribute to the high complexity of traditional PKIs. 

\subsubsection*{Private key management}

The choice of a private key container dictates many characteristics of a PKI and how users interact with it. Hardware devices struggle with affordability and interoperability; laypeople are commonly unable to obtain or interface directly with HSMs, and although TPMs may be certified by CAs, in practice, they are commonly associated with \emph{devices} instead of end users. In the case of smartcards and tokens, devices may only work with a specific reader or operating system version. Such devices may also be stolen or broken. As they hold private keys, the owner must always notify the CA of any complications so that it can revoke the corresponding digital certificate~\cite{Boeyen:200805}. Moreover, while relatively cheap when compared to HSMs, the acquisition of these devices for each user in a widespread PKI environment may hinder its deployment, especially in countries with limited resources.

Conversely, cloud-based solutions have been essential to the growth of digital signatures for the general public~\cite{Chong:202101} and show that digital signature adoption can be increased by relieving the user from the burden of carrying a physical device. However, several issues have already been raised in the literature~\cite{Damgard:201312,Padilha:201501,Kumar:202010}. Signers no longer have unique possession of their keys, as they are now stored and possibly replicated in cloud servers. Furthermore, the inherent lack of confidentiality of cloud storage diminishes the assurance that a user has that its private key has not been used by any other entity. Supporters of cloud-based solutions claim that there are KMaaS tools and procedures that allow users to audit and verify the authorized usage of their cryptographic keys. 

Alas, these requirements often neglect that a widespread PKI model would have laypeople as its main users and, therefore, cannot expect them to perform these technical operations. We observe that the issues brought forward by the need for a secure key management model are inherent to the combined existence of the private key and a layperson end user. Therefore, any model with these characteristics is subject to some form of a key management challenge. We argue that, to address these issues while designing the model, one must (i) restrict the type of user allowed to participate or (ii) remove the need for a private key container entirely. We choose the latter, as we aim to define an inclusive, user-friendly PKI.

\subsubsection*{Certificate revocation}

Revocation is probably the most complex and vulnerable service for CAs, registration authorities (RAs), and users of a traditional PKI. In the literature, criticisms about this mechanism are regular~\cite{Zheng:200304,Kocher:199802,Raya:2006,Wang:202001,Naor:199801,Bour:202011}, and there exist several alternative methods to facilitate, speed up and guarantee the verification of the status of a digital certificate. However, Årnes~\cite{Arnes:200002} surveys most of these revocation mechanisms, and their conclusion is that they often present several shortcomings and narrow suitability, with no definitive solution yet known. A common denominator is that private key compromises must be preemptively detected before they can be addressed.
 
We present a non-exhaustive list of issues with the CRL model, as follows: (i) lists must be issued even if no certificates are revoked; (ii) compromised private keys can be used until the newly published list contains the respective revoked certificate; (iii) lists can grow to large file sizes, which can congest network channels and quickly fill non-volatile storage media; (iv) network load spikes happen when a list expires~\cite{Cooper:199912}; (v) lists may become inaccessible due to server outages; and (vi) human error can cause certificates to be revoked incorrectly.

OCSP also has its shortcomings~\cite{Gutmann:200208,Wohlmacher:200011,Santesson:201306}. Queries are online and thus increase the load on communication networks. Furthermore, servers are subject to denial of service attacks. OCSP servers may provide fresher revocation information when compared to CRLs. However, many OCSP services do not have direct access to CA databases due to security concerns and use CRLs as a primary source of revocation data, thus inheriting the latency issues of CRLs. Finally, querying revocation information may indirectly reveal sensitive data about the petitioner.

As previously mentioned, revocation is essential to maintaining trust in the PKI because private keys are always subject to compromises, and attributes in certificates become outdated. We argue that a model that minimizes such threats is able to reduce the consequences of maintaining and providing revocation information. End users comprise most certificate holders in a PKI and are commonly liable for having certificates revoked. Any model that aims to reduce the burden of revocation must target such users. Therefore, we propose a PKI framework that provides irrevocable certificates to end users.

\section{One-time digital certificates}\label{sec:otc}

We propose a new cryptographic key management model to address the aforementioned issues. The main differences between our model and the traditional PKI are threefold: (i) we bind a digital certificate to a single document, preventing loss of authenticity from private key compromises; (ii) we reduce the lifecycle of end-user private keys, lowering the complexity associated with their management; and (iii) we forego revocation mechanisms at the end-user level. We discuss those traits below and compare them to the traditional model.

After key pair generation, the public key must be certified by a CA and bound to a certificate. This process does not change between a classic PKI and our proposal. However, to ensure that signed documents are validated correctly for long periods of time, we propose to bind a document to a certificate via a signed attribute containing the cryptographic hash of the original document. To validate a digitally signed document in our model, in addition to the traditional verification steps, the digest of the original document is matched against the digest embedded in the certificate.

\begin{figure}[htbp]
  \centering
  \begin{tikzpicture}[x=0.75pt, y=0.75pt, yscale=-1]
    \draw[-Latex] (36, 12) -- (36, 0) -- (264, 0) node [midway, font=\scriptsize] {\Circled{1}} -- (264, 12);
    \draw[-Square, purple] (192, 48) -- (264, 48) node [midway, font=\scriptsize] {\Circled[fill color=purple]{2}};
    \draw[-Latex] (264, 36) -- (264, 88) node [pos=0.6, font=\scriptsize] {\Circled{3}};
    \draw[-Latex] (240, 100) -- (216, 100) node [midway, font=\scriptsize] {\Circled{4}} -- (216, 84) -- (192, 84);
    \draw[-Square] (96, 48) -- (36, 48) node [midway,  font=\scriptsize] {\Circled{5}};
    \draw[-Latex] (36, 36) -- (36, 88) node [pos=0.6, font=\scriptsize] {\Circled{6}};
    \draw[-Latex] (36, 112) -- (36, 120) -- (105, 120) node [midway, font=\scriptsize] {\Circled{7}};
    \draw[rounded corners=4pt] (0, 12) rectangle (72, 36) node [midway, black] {Signer};
    \draw[rounded corners=4pt] (228, 12) rectangle (300, 36) node [midway, black] {CA};
    \draw[orange] (108, 18) -- (108, 66) -- node [midway, above, black, yshift=10pt] {Document $(\mathsf{d})$}(192, 66) [rounded corners=4pt] -- (192, 18) -- cycle;
    \draw[orange] (108, 72) rectangle (192, 96) node [midway, black] {Certificate $(\mathsf{c})$};
    \draw[dashed, rounded corners=4pt] (96, 12) rectangle (204, 102);
    \draw (12, 88) rectangle (60, 112) node [midway, black] {\textsc{Sig}};
    \draw (240, 88) rectangle (288, 112) node [midway, purple] {OTC};
    \draw[orange] (192, 132) -- (192, 108) -- (108, 108) [rounded corners=4pt] -- (108, 132) -- cycle node [midway, above, black] {Signature $(\sigma)$};
  \end{tikzpicture}
  \caption{
    Ordered steps to create a digitally signed document using the OTC model. We assume that a key pair has been previously generated by the signer.
    In step \Circled{1}, the public key is sent by the signer to the CA.
    In step \Circled[fill color=purple, outer color=white]{2}, the signer computes the hash of the original document and sends it to the CA.
    In step \Circled{3}, the CA issues an OTC with fresh attributes to the signer.
    In step \Circled{4}, the signer appends their OTC to the original document.
    In step \Circled{5}, the signer computes the hash $\mathsf{h}$ of the tuple $(\mathsf{d}, \mathsf{c})$.
    In step \Circled{6}, the signer signs $\mathsf{h}$ with their algorithm of choice.
    In step \Circled{7}, the signature is appended to the document, creating the final artifact $(\mathsf{d}, \mathsf{c}, \sigma)$ in \textcolor{orange}{orange}.
  }\label{fig:otc-signing}
\end{figure}
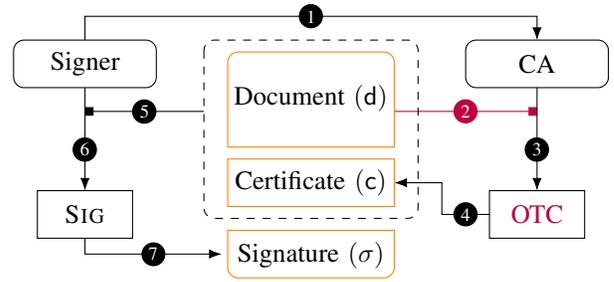

The signing process of a document under the \emph{one-time certificate} (OTC) model is exemplified in Figure~\ref{fig:otc-signing}. We observe that most steps and artifacts follow the traditional procedures for generating a signed digital document. The only change is the addition of step \Circled[fill color=purple, outer color=white]{2}: the embedding of the document hash in the certificate along with the user identity information, effectively creating a unique binding between user, key pair, and document. 

In our model, while a private key can be used to generate any number of signatures, only the document identified in the corresponding certificate will be correctly validated. Thus, a private key associated only with an OTC has no further \emph{effective} use and may be safely \emph{destroyed} after document signing. Consequently, we are able to use unique key pairs for each signature generation procedure. The private key lifecycle is not abruptly interrupted by compromises due to the aforementioned unique binding. In other words, a leaked private key may only be used to sign a specific document. Furthermore, document signature and private key deletion do not need to be atomic operations, as the OTC binding prevents effective key misuse even if the key is leaked between these operations. 

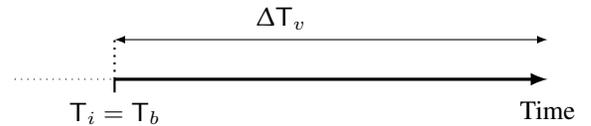
\begin{figure}[htbp]
  \centering
  \begin{tikzpicture}[very thick, node distance = 4cm] 
    \draw[dotted, thin] (0, 0) -- (0.15\linewidth, 0);
    \draw[latex-latex, thin] (0.15\linewidth, 15pt) -- (0.8\linewidth, 15pt);
    \draw[-latex] (0.15\linewidth, 0) -- (0.8\linewidth, 0);
    
    \draw[thick] (0.8\linewidth, -4.5pt) node[below, align=center] {Time};
    \draw[thick] (0.4\linewidth, 15pt) node[above, align=center] {$\DTc{v}$};
    \draw[thick] (0.15\linewidth, -5pt) node[below, align=center] {$\Tc{i}=\Tc{b}$};
    
    \draw [thick] (0.15\linewidth, 0) -- (0.15\linewidth, -5pt);
    \draw [dotted, thick] (0.15\linewidth, 0) -- (0.15\linewidth, 15pt);
  \end{tikzpicture}
  \caption{Time references for an OTC.}\label{fig:reference_time_otc}
\end{figure}

A natural consequence of the unique binding between signer, certificate, and document is that the underlying public key must be certified immediately before every signature generation. As the tuple of signer, digital certificate and document is unique, issuance happens whenever a signature must be generated. As previously mentioned, it is the responsibility of the CA to ensure that attributes are not stale throughout the lifetime of the certificate. Therefore, when a certificate is issued, every attribute is valid. In our model, since certificate issuance happens immediately before signature generation, all attributes can be assumed valid at the time of signature generation $\Tc{\sigma}$. Therefore, timestamps are not required when using OTCs. Figure~\ref{fig:reference_time_otc} shows the OTC lifetime.

Consequently, by removing the eventual staleness of the attributes and preventing private key compromises due to a short lifecycle derived from the unique binding between certificates and documents, we argue that revocation mechanisms at the end user level are unnecessary. In short, we remove the association between the lifecycle of the private key and the validity period of the corresponding digital certificate. We note that OTCs are \emph{not ephemeral} since they are valid and may be verified for the lifecycle of the signature, which is only bound by the cryptographic strength of the underlying algorithm.

\subsection{Specifications \& interoperability}

Our proposal is technically simple and can be put into practice with minor overhead. It is backward-compatible with current PKI standards, such as the well-known RFC 5280 profile~\cite{Boeyen:200805}, signature validation applications (SVAs), and conventional cryptographic algorithms. One-time certificates are pursuant to the X.509 profile, considering the following characteristics. The \texttt{notAfter} entry of the \texttt{Validity} ASN.1 structure inside \texttt{TBSCertificate}, which represents $\Tc{a}$ in Figure~\ref{fig:reference_time}, is set to an arbitrarily remote date in the future, beyond the intended lifetime of any digitally signed document.

OTCs do not require any revocation mechanisms, such as CRLs or OCSP queries. However, there must exist such revocation information available in specific X.509 certificate extensions, either through \texttt{id-ce-cRLDistributionPoints}~\cite[Sec.~4.2.1.13]{Boeyen:200805} in the case of CRLs or \texttt{id-pe-authorityInfoAccess}~\cite[Sec.~4.2.2.1]{Boeyen:200805} for OCSP servers. Therefore, a CA that issues OTCs should also issue a single, empty CRL. The revocation list must inherit the validity period of the CA. Alternatively, an OTC-compliant OCSP server may be deployed. OCSP responses must return a \texttt{CertStatus} entry of \texttt{good} for any queries about OTCs issued by CAs it is responsible for or \texttt{unknown} for queries about other certificates.

One-time certificates must contain an attribute representing the binding between the user, the certificate, and the to-be-signed document. It is a critical certificate extension, i.e., an ASN.1 structure of type \texttt{Extension}, whose \texttt{extnValue} is of type \texttt{OneTimeCertificateHash}, according to the syntax below, pursuant to RFC 5912~\cite{Hoffman:201006}. The object identifier \texttt{extnID} of the extension is yet to be defined in future standards. The choice of the cryptographic hash algorithm is collected from the end user by the signature creation application (SCA).
\begin{verbatim}
OneTimeCertificateHash ::= SEQUENCE {
  hashAlgorithm       AlgorithmIdentifier,
  documentHash        OCTET STRING
}
\end{verbatim}

As previously mentioned, the binding between a document and an end-user through the certificate is a major aspect that grants OTCs the ability to waive revocation and secure key pair generation. Thus, SVAs \emph{must} validate this binding in the form of a comparison between the contents of the \texttt{OneTimeCertificateHash} extension and the digest of the original document. Otherwise, legacy SVAs may be tricked into accepting electronic documents with signatures that are correctly \emph{verified} by the public key in the OTC but incorrectly \emph{validated} due to mismatching digests. We recall that SVAs are required to reject certificates with unknown critical extensions. Thus, by setting the \texttt{OneTimeCertificateHash} to critical, we prevent any compliant legacy SVAs from being tricked.

\subsection{Certificate issuance \& authentication}

While a thorough analysis of different authentication methods is out of the scope of this work, as OTCs function independently from the registration and authentication methods used by the infrastructure, automating the issuance process after a user has been registered greatly increases the usability of OTCs and the PKI in general. Nonetheless, there are several adjustments that can be made in the registration and authentication steps of the issuance process to fit any particular environment. Smaller PKIs might still opt for manual registration and authentication, local PKIs may opt for manual registration and automatic certificate issuance through an identity provider (IdP), and a high-load widespread PKI is able to offer both manual and automated registration and certificate issuance with different levels of assurance.

We assume the user has already been registered in the IdP by whatever methods are required by the PKI. An example of the routine performed by a SCA is given as follows: (i) the user accesses its SCA; (ii) the SCA redirects the user to a trusted IdP; (iii) the user authenticates itself and allows the sharing of its information; (iv) the user submits the document to be signed; (v) the SCA generates a key pair, certificate signing request, the hash of the document and forwards them to the CA along with the signed IdP token; (vi) the CA issues a certificate containing up-to-date user information and the hash of the document and returns it to the SCA; and (vii) the SCA generates the signature according to the chosen policy and returns the signed document to the user.


We observe that a direct consequence of requiring frequent authorization from an IdP leads to a stronger assurance that the original holder is the one who signed a document when compared to traditional models. That is, traditional PKIs make several assumptions regarding users and their key management routines. One such assumption refers to the use of the private key, which should only be used by the original holder of the certificate, i.e., ``lending'' a private key should not happen. Alas, such assumptions are hardly realistic as there is little guarantee of who is using the private key at a given moment.

Conversely, under the OTC model, users are required to authenticate themselves to sign new documents, such that a compliant PKI may employ IdPs with several different mechanisms to reliably ascertain that a given user is accessing the system. IdPs may require users to offer sufficient proof that they are the ones authenticating themselves, such as through multi-factor authentication methods, enhancing the assurance that the original holder is the one signing the document.

\subsection{Privacy considerations}

In the traditional model, users are able to acquire certificates and sign any number of documents without revealing any information about the documents they sign. However, the inclusion of the hash of the document in the certificate in the OTC model reveals some, albeit minimal, information about the user and the document it wishes to sign. In this manner, CAs could track high-profile public documents and refuse to issue certificates for users who wish to sign them. 

PKIs that are worried about the possibility of this denial of service attack for specific documents are able to circumvent it by including a \emph{nonce} in the signature, appending it to the original document before generating the hash to be included in the certificate. SVAs that detect the presence of this nonce must append it to the document before generating the hash to be validated with the certificate. In this manner, offending CAs are not able to track public documents and refuse to issue certificates in a document-by-document case.

\subsection{Security analysis}

We argue that our proposal is at least as secure as the traditional PKI model, as it restricts the number of documents that can be certified by a digital certificate. We emphasize that threats arising from breaches of trusted parties, such as CAs and RAs, are not in the scope of our proposal. These issues are commonly addressed via the Certificate Transparency~\cite{Laurie:202112} framework. Hereafter, we show that OTCs solve the problem of stale attributes and private key management, eliminating the need for revocation mechanisms at the end-user level.

\begin{definition}\label{def:staleness}
  An attribute is considered \emph{stale} or \emph{outdated} if it is embedded within the digital certificate and does not correctly identify the entity when a signature is generated.
\end{definition}

\begin{proposition}\label{proposition:attributes}
  OTC attributes cannot be stale.
\end{proposition}
\begin{proof}
  Let $\mathsf{pk}$ be a public key and $\mathsf{c}$ its corresponding OTC. We recall that $\Tc{i}$ is the moment when $\mathsf{c}$ is created from $\mathsf{pk}$ and attributes, and $\Tc{b}$ is the start of its validity period. Such attributes are valid at $\Tc{i}$ due to the role of a CA. In our model, the validity period of $\mathsf{c}$ starts immediately after issuance, such that $\Tc{i} \lesssim \Tc{b}$. Therefore, the attributes contained in $\mathsf{c}$ and validated at $\Tc{b}$ are guaranteed to be accurate.
\end{proof}

A consequence of Proposition~\ref{proposition:attributes} is that certificate issuance acts as a time-mark, and document binding prevents certificate reuse. Therefore, signatures under instances of our model do not require timestamps to achieve the same characteristics of other policies. Additionally, we note that any changes to user attributes would be reflected in new OTCs. Furthermore, consider $\Tc{\sigma}$ to be the moment a signature $\sigma$ is generated with $\mathsf{sk}$. While we expect $\Tc{b} \lesssim \Tc{\sigma}$, it is not a prerequisite since attributes are validated at $\Tc{b}$.

\begin{definition}[\cite{bellare1999forward}]\label{def:forward}
The \emph{forward security} property indicates that it is computationally impractical for any attacker to forge a valid signature for a time period before the compromise of the corresponding private key. 
\end{definition}

Traditionally, forward security proofs include a forger $\mathsf{F}$ that conducts an adaptive chosen message attack~\cite{Goldwasser:1988} on requested messages of its choice until it chooses to ``break-in'' at time $\Tc{b}$ by requesting the private key $\mathsf{sk}$ and generating a distinct signed message $\mathsf{m}$ at time $\Tc{\sigma}$ such that $\Tc{\sigma} < \Tc{b}$; that is, the forged signature must be generated for a time before its compromise. The forgery is successful if the signature is considered valid~\cite{itkis2001forward}.

We extend this traditional notion of forward security to include certificate policies and show that documents signed using the OTC model are forward-secure. In our model, the forger $\mathsf{F}$ makes an adaptive chosen message attack on requested documents receiving the corresponding signed document $\sigma_i$ and OTC $\mathsf{o_i}$ until it requests the secret key $\mathsf{sk}$ at time $\Tc{b}$; the forgery is considered valid if the forger is able to sign a distinct document $\mathsf{d}$ and successfully validate it using any of the issued OTC $\mathsf{c}$ under the correct policy for any time before $\Tc{b}$.

\begin{proposition}\label{proposition:forward}
    The OTC model is forward-secure.
\end{proposition}
\begin{proof}
Let $\mathcal{D}$ be the set of documents requested by the forger, $\mathcal{S}$ the set of signatures, and $\mathcal{O}$ the set of OTCs received by the forger such that $\mathsf{d}_{i}$ is bound to $\mathsf{c}_{i}$ and the tuple $(\sigma_i \in \mathcal{S}, \mathsf{d}_{i} \in \mathcal{D}, \mathsf{c}_{i} \in \mathcal{O})$ is valid. The forgery is considered a success if the forger is able to generate a valid tuple $(\sigma_i, \mathsf{d}_{f}, \mathsf{c}_{i})$ such that $\mathsf{d}_{f} \neq \mathsf{d}_{i}$. However, as a consequence of the OTC binding, for $(\sigma_i, \mathsf{d}_{f}, \mathsf{c}_{i})$ to be valid, $\mathsf{d}_{f} = \mathsf{d}_{i}$.
\end{proof}

The threat modeling for the forward security property can be expanded to include \emph{any} signatures made by the forger, not only those before $\Tc{b}$. The removal of this constraint from the attacker leads to the stronger security notion of \emph{unforgeability}~\cite{Goldwasser:1988}; that is, no signature made by the forger is considered valid. Extending this concept for forward security with certificate policies is straightforward, requiring the removal of the same constraint, and shows that a signed document is unforgeable under a policy.

We note that the proof for Proposition~\ref{def:forward} does not depend on any time frames to be correct, neither from the forger or the oracle; the forger may generate a forgery for any time, and the forgery would still be unsuccessful as a result of the restrictions imposed by the OTC model and its binding. We show that restricting the number of valid verifications to a single document per digital certificate hinders forgeries in such a manner that not only is our model forward secure, but also that signed documents associated with OTCs are unforgeable.

\begin{proposition}\label{proposition:oneuse}
    Signatures linked to OTCs are unforgeable.
\end{proposition}
\begin{proof}
  Assume trustworthy issuer CAs, a secure cryptographic hash function, and a compliant signature verification application that correctly validates the digest of the signed document against the embedded hash. Let $(\mathsf{sk}, \mathsf{pk})$ be a key pair associated only with OTCs, an OTC $\mathsf{c}$ that certifies $\mathsf{pk}$ and a document $\mathsf{d}$, and $\mathcal{S}$ to be the set of signatures signed by $\mathsf{sk}$ that are correctly validated with $\mathsf{c}$. We recall that forgeries must differ from existing signatures and thus require that $|\mathcal{S}| > 1$. However, the OTC model shrinks $\mathcal{S}$ to one signature. Therefore, as $|\mathcal{S}| = 1$ under the OTC model, any other signatures generated with $\mathsf{sk}$ fail to validate using $\mathsf{c}$; these signatures cannot be forged.
\end{proof}

We note from Proposition~\ref{proposition:oneuse} that a private key associated only with OTCs cannot be \emph{effectively} compromised by any outside attacker; that is, an attacker with access to the user private key cannot generate new valid signatures. Furthermore, Proposition~\ref{proposition:oneuse} allows the relaxation of the security requirements for the user signature, i.e., as long as the signature algorithm and the private key used by the CA are secure, the breaking of the algorithm used by the user would not invalidate previously issued certificates and signed certificates.

This property allows key pair generation not to require secure private containers, which usually have lower key generation throughput. Hence, private keys can be generated where they are most convenient, in environments such as embedded devices, cloud services, and personal computers, improving usability. We note that key pair reuse is not forbidden and does not hinder the security of our model due to the unique binding between a certificate and a document. However, to keep usability in the foreground, we assume key pairs are not stored and thus must be obtained immediately prior to a signature.

Furthermore, Proposition~\ref{proposition:oneuse} also enables the certification of one-time signatures under the traditional model of CSR generation before certificate issuance. Traditionally, the second use of the OTS private key would reveal information about the private key. However, under the OTC model, no other signatures can be valid, negating the issues of leaking private key information of OTS schemes. We highlight that our proposal is crypto-agile and does not require the use of OTS schemes, but it does not prevent their use.

Nevertheless, it is worth noting that while OTCs protect the end user from forgeries, it does not solve traditional security issues resulting from corrupt or negligent issuer CAs; that is, issuer CAs can still generate valid certificates for any given combination of user and document without being given explicit consent from the user. Furthermore, we remark that the security of OTCs largely depends on the security of the hash function used to bind the document to the certificate; we recommend the use of hash functions that are at least as secure as the ones used in the signature process.

\section{Performance evaluations}\label{sec:performance}

Generating a key pair and issuing a new digital certificate before signature generation has heavy implications for the performance of creating digitally signed electronic documents. In this section, we measure the overhead caused by generating key pairs on demand, discuss its consequences, and propose alternative key generation methods to evade such performance costs. We enumerate how many signatures are needed to successfully sign and validate a document using traditional signature policies and OTCs. Finally, we calculate the expected throughput of document signing and validation and present our deployment recommendations considering several use cases.

In the traditional PKI model, an end user is expected to possess a certified key pair already. Consequently, only the performance of operations on signature is a relevant metric. However, when assessing the performance of OTCs, we must also consider the performance of key generation when estimating the number of signed documents an end user is able to create and validate with no detriment to their workflow.

We first measure the performance of several digital signature schemes already used in PKIs for various security levels based on the estimations in~\cite[Table~2]{Barker:202005}. Hereafter, our measurements are performed in a GNU/Linux machine with kernel version 6.5.7, equipped with an AMD Ryzen™ 5700G @ 3.8GHz, ``turbo boost'' disabled, and OpenSSL 3.1.3. All operations on signatures use the SHA-256 cryptographic hash function when applicable. Table~\ref{tab:algorithms} shows that requiring key pair generation before every document signature introduces a bottleneck in the document signature process. Thus, it is clear that key generation speed becomes a critical performance component for OTCs. 

\begin{table}[htbp]
  \centering
  \setlength{\tabcolsep}{6pt}
  \caption{Average performance of $2^{10}$ iterations, in microseconds, for each operation of several commonly used digital signature algorithms with various security levels $n$.}\label{tab:algorithms}
  \begin{tabular}{lrrrrr}
    \toprule
    &
      \multicolumn{3}{c}{$n = 128$} &
      \multicolumn{1}{c}{$n = 192$} &
      \multicolumn{1}{c}{$n = 224$} \\
    \cmidrule(rl){2-4} \cmidrule(rl){5-5} \cmidrule(rl){6-6}
    &
      \multicolumn{1}{c}{RSA-3072} &
      \multicolumn{1}{c}{P-256} &
      \multicolumn{1}{c}{Ed25519} &
      \multicolumn{1}{c}{P-384} &
      \multicolumn{1}{c}{Ed448} \\
    \midrule
    \textsc{KeyGen}  & 122864.62  & 21.04  & 34.91   & 612.64  & 176.16  \\
    \textsc{Sig}  & 1448.06    & 17.17  & 30.78   & 636.99  & 173.60  \\
    \textsc{Ver}  & 29.84      & 49.32  & 84.14   & 530.75  & 194.34  \\
    \bottomrule
  \end{tabular}
\end{table}

We present two methods to address this trade-off, loosely referred to as policies. The first is the generation of key pairs on demand, hereafter called \textsf{OTC-D}. This is the model presented in Section~\ref{sec:otc}, which allows us to omit key management. We recall that Proposition~\ref{proposition:oneuse} removes the potential of private key compromises. Therefore, private keys and CSRs can be generated in advance and stored for later use. We call this method of pre-generating artifacts \emph{bulk key generation} or \textsf{OTC-B}. This method introduces some light key management but manages to avoid most of the overhead introduced by our model, i.e., the on-demand generation of key pairs. We note that these CSRs are built only to show private key ownership and do not require user identity information or the document hash; these are sent to the CA when requesting a new certificate.

To sign a document using \textsf{OTC-B}, the signer requests a pre-generated CSR, requests an OTC, signs the document, and discards the private key. Furthermore, in both methods, key generation and storage can be delegated to other entities, allowing low-power devices to sign documents using OTCs in a timely manner. However, we note that neither \textsf{OTC-B} nor \textsf{OTC-D} prevent the overhead of repeated certificate issuance at the CA level; this is an unavoidable trade-off of our model.

\subsection{Comparison to traditional signature policies}

We recall that signature policies can augment a digitally signed document with further signed artifacts, such as revocation information and timestamps. In this section, we discuss how OTCs can be used in the context of Advanced Electronic Signatures. We compare the number of signatures required by European policies~\cite[Table~1]{ETSI-EN-3191221-121:202110} to process digitally signed documents to semantically equivalent documents created with OTCs. Finally, we use the measurements of Table~\ref{tab:algorithms} to estimate the performance of our model in practical use cases.

Timestamps greatly affect the performance of processing digitally signed documents, as seen in Section~\ref{sec:ades}. Distinct policies require a different number of timestamps~\cite[Table~1]{ETSI-EN-3191021-131:202111}. The set of \textsf{B} policies does not require timestamps and requires the least amount of information to be created; \textsf{T}-type signatures include a time-mark in the form of a timestamp to provide information about when the signature was created; policies of type \textsf{LT} additionally include certificates from the whole chain to the set of signed artifacts of \textsf{T} signatures; the archival signature policy \textsf{LTA} includes an additional timestamp and complete certification and revocation artifacts to the signature. Furthermore, every policy requires the inclusion of the user signing certificate as a signed attribute. 

We recall from Proposition~\ref{proposition:attributes} that OTCs act as time-marks due to their issuance procedures. Let $\varsigma$ be the minimal amount of signatures a document carries when created according to a policy $p$. We split this number into quantities representing how many signatures are created for every step in the PKI model: (i) $\sigma_{d}$ is the number of signatures that cover the original document; (ii) $\sigma_{c}$ is the number of signatures needed in a CSR; (iii) $\sigma_{i}$ is the number of signatures needed to issue a certificate; and (iv) $\sigma_{t}$ is the number of timestamps required. The number of operations required by each policy is shown in Table~\ref{tab:signature_ops}.

\begin{table}[ht]
  \centering
  \setlength{\tabcolsep}{7.5pt}
  \caption{Number of key and signature operations required by a policy $p$ in the lifecycle of a digitally signed electronic document for a certificate chain of depth $c$.} \label{tab:signature_ops}
  \begin{tabular}{lrrrrrrrr}
    \toprule
    \multicolumn{1}{c}{\multirow{2}{*}{$p$}} &
      \multicolumn{1}{c}{\multirow{2}{*}{\textsc{Gen}}} &
      \multicolumn{5}{c}{\textsc{Sig}} &
      \multicolumn{2}{c}{\textsc{Ver}} \\
    \cmidrule(rl){3-7} \cmidrule(rl){8-9}
    \multicolumn{1}{c}{} &
      \multicolumn{1}{c}{} &
      \multicolumn{1}{c}{$\sigma_{d}$} &
      \multicolumn{1}{c}{$\sigma_{c}$} &
      \multicolumn{1}{c}{$\sigma_{i}$} &
      \multicolumn{1}{c}{$\sigma_{t}$} &
      \multicolumn{1}{c}{$\varsigma$} &
      \multicolumn{1}{c}{$c = 3$} &
      \multicolumn{1}{c}{$c = 4$} \\
    \midrule
    \textsf{B}     & \textbf{0} & 1 & 0 & 0 & 0 & \textbf{1} &  \textbf{6} & \textbf{8}  \\
    \textsf{T}     & \textbf{0} & 1 & 0 & 0 & 1 & \textbf{2} & \textbf{12} & \textbf{16} \\
    \textsf{LT}    & \textbf{0} & 1 & 0 & 0 & 1 & \textbf{2} & \textbf{12} & \textbf{16} \\
    \textsf{LTA}   & \textbf{0} & 1 & 0 & 0 & 2 & \textbf{3} & \textbf{18} & \textbf{24} \\
    \midrule
    \textsf{OTC-B} & \textbf{0} & 1 & 0 & 1 & 0 & \textbf{2} &  \textbf{5} & \textbf{7}  \\
    \textsf{OTC-D} & \textbf{1} & 1 & 1 & 1 & 0 & \textbf{3} &  \textbf{5} & \textbf{7}  \\
    \bottomrule
  \end{tabular}
\end{table}

We note that the removal of timestamps in the OTC model lowers the amount of signatures required to be validated down to $s = 2c$. Additionally, SVAs compatible with our proposal need not support revocation processing at the signer level according to the certification path validation algorithm of RFC 5280~\cite[Sec.~6.3]{Boeyen:200805}. Hence, one less signature must be verified throughout the algorithm: the empty CRL or OCSP response, further reducing the number of validations to $s = 2c - 1$.

Finally, we present an analytical throughput measurement for each policy and key generation method and give our deployment recommendations based on the intrinsic characteristics of our model. Table~\ref{tab:throughput} estimates how many digitally signed documents can be created and validated per second by using the algorithm measurements of Table~\ref{tab:algorithms} and the number of operations per policy in Table~\ref{tab:signature_ops}.

\begin{table}[htbp]
  \centering
  \setlength{\tabcolsep}{3pt}
  \caption{Throughput measurements, in \textbf{documents per second}, of the creation and validation of digitally signed documents with a policy $p$ and $c = 4$, considering commonly used digital signature algorithms with various security levels $n$.}\label{tab:throughput}
  \scriptsize
  \begin{tabular}{lrrrrrrrrrrrr}
    \toprule
    \multicolumn{1}{c}{\multirow{3}{*}{$p$}} &
      \multicolumn{4}{c}{$n = 128$} &
      \multicolumn{2}{c}{$n = 192$} &
      \multicolumn{2}{c}{$n = 224$} \\
      \cmidrule(rl){2-5} \cmidrule(lr){6-7} \cmidrule(lr){8-9}
    & 
      \multicolumn{2}{c}{RSA-3072} &
      \multicolumn{2}{c}{Ed25519} &
      \multicolumn{2}{c}{P-384} &
      \multicolumn{2}{c}{Ed448} \\
      \cmidrule(rl){2-3} \cmidrule(lr){4-5} \cmidrule(lr){6-7} \cmidrule(lr){8-9}
    &
      \multicolumn{1}{c}{\textsc{Sig}} &
      \multicolumn{1}{c}{\textsc{Ver}} &
      \multicolumn{1}{c}{\textsc{Sig}} &
      \multicolumn{1}{c}{\textsc{Ver}} &
      \multicolumn{1}{c}{\textsc{Sig}} &
      \multicolumn{1}{c}{\textsc{Ver}} &
      \multicolumn{1}{c}{\textsc{Sig}} &
      \multicolumn{1}{c}{\textsc{Ver}} \\
    \midrule
    \textsf{B}      &  690.58   &  4188.76  &  32484.22  &  1485.57  &  1569.89  &  235.07  &  5760.25  &  643.21  \\
    \textsf{T}      &  345.29   &  2094.38  &  16242.11  &  742.79   &  784.94   &  117.54  &  2880.13  &  321.61  \\
    \textsf{LT}     &  345.29   &  2094.38  &  16242.11  &  742.79   &  784.94   &  117.54  &  2880.13  &  321.61  \\
    \textsf{LTA}    &  230.19   &  1396.25  &  10828.07  &  495.19   &  523.30   &  78.36   &  1920.08  &  214.40  \\
    \midrule
    \textsf{OTC-B}  &  345.29   &  4787.15  &  16242.11  &  1697.80  &  784.94   &  268.65  &  2880.13  &  735.10  \\
    \textsf{OTC-D}  &  2.68     &  4787.15  &  5073.85   &  1697.80  &  266.75   &  268.65  &  953.02   &  735.10  \\
    \bottomrule
  \end{tabular}
\end{table}

As expected, creating signed documents under \textsf{OTC-D} is hindered by the performance of key generation, especially notable when using RSA. On the other hand, \textsf{OTC-B} offers competitive performance compared to all policies that enforce timestamps. We note that both OTC policies excel at the document validation step since there are fewer certificate chains and revocation artifacts to be verified. This is highly sought after; documents are more often validated than digitally signed.

Ultimately, each application must decide on the acceptable throughput for their systems and which key generation mechanism better suits its purpose. Applications that sign documents once every few minutes might want to choose the simpler on-demand method. At the same time, systems with low processing power might need to delegate this task to a server, and applications that require high throughput may pre-generate key pairs. Systems may distribute the load between signature creation and validation routines, and stakeholders may provide only one of the two services. Our model does not hinder such adjustments. 

Furthermore, we remark that the load increase to issuer CAs caused by our model is comparable to the standard load of KMaaS applications; we recall that such applications must activate and make use of an end-user private key to generate a signature for each document. Likewise, an OTC issuer CA must employ its own private key to issue new OTCs.

The benefits of using OTCs cannot be overstated. Revocation at the end-user level has been removed, private key containers are no longer a concern, and signature policies have been made much simpler, particularly through the redundancy of timestamps. As a whole, the creation and validation of digitally signed documents is simplified at the expense of lower signature generation performance. However, it can be tweaked to desirable levels using specific key generation strategies and different digital signature algorithms.

\section{Conclusion}\label{sec:conclusion}

Private key management and certificate revocation have been sources of complexity in PKIs since their inception. In this work, we introduced a new key management model capable of solving these issues at the end-user level through the unique binding of a digital certificate and a document using our one-time certificates. We show that many artifacts responsible for additional complexity in PKIs are no longer needed in our proposal, and security challenges associated with private key compromises and stale attributes are prevented in our model.

We maintain usual signature policy requirements when signing documents using OTCs while greatly improving the performance of validation procedures simultaneously. Trade-offs between usability and the performance of document creation can be lessened through clever algorithm choices and key generation strategies. These features enable the creation of a simple, efficient, and inclusive PKI, built for the rapid adoption of digitally signed electronic documents by laypeople, with several usability features and robust security properties.

\subsubsection*{Future works}

The unique binding of a certificate and document additionally enables the monitoring of the creation of digitally signed documents. A certificate misissuance detection protocol, analogous to the Certificate Transparency~\cite{Laurie:202112} framework, can be developed from our proposal. The behavior of our model with quantum-safe signature schemes and hybrid certificates can also be explored; we anticipate that the size of public keys and signatures and the performance of operations on signatures will present unique challenges.

Finally, large communities, such as countries with populations in the tens of millions, might see a sizable cost reduction in deploying and providing affordable digital signatures of electronic documents to citizens. These savings come mainly from the inessential provision of private key containers for the populace and less management overhead at the final CA level due to revocation-less certificates.
z




\bibliographystyle{amsplain}
\bibliography{ref}

\end{document}